\documentclass[oneside,english]{article}
\usepackage[a4paper]{geometry}

\usepackage[activate={true,nocompatibility},final,tracking=true,kerning=true,spacing=true,factor=1100,stretch=10,shrink=10]{microtype}

\usepackage[english]{babel}
\usepackage[numbers,sort]{natbib}

\usepackage[utf8]{inputenc}
\usepackage[T1]{fontenc}
\usepackage{lmodern}

\usepackage[dvipsnames]{xcolor}
\usepackage{xcolor-grey}

\usepackage{amsmath}
\usepackage{amsfonts}
\usepackage{amssymb}
\usepackage{amsthm}
\usepackage{mathtools}

\usepackage{subcaption}

\usepackage{hyperref}
\usepackage{nameref}

\newtheorem{observation}{Observation}

\theoremstyle{definition}
\newtheorem{example}{Example}

\addto\extrasenglish{

}

\addto\extrasenglish{

}

\usepackage{enumerate}
\usepackage[shortlabels,inline]{enumitem}

\usepackage{booktabs}
\usepackage{tabularx}
\usepackage{makecell}

\usepackage{url}

\newcommand{\ie}{i.\nolinebreak[4]\nolinebreak[4]e.}

\newcommand{\lO}{\mathcal{O}}
\newcommand{\lo}{o}

\newcommand{\setB}{\mathbb{B}}
\newcommand{\setN}{\mathbb{N}}

\newcommand{\occ}{\text{occ}}
\newcommand{\ropenint}[2]{[#1,#2)}
\newcommand{\lsceil}{\lceil\log\sigma\rceil}

\newcommand{\eff}{\text{eff}}
\newcommand{\rank}{\text{rank}}
\newcommand{\select}{\text{select}}
\newcommand{\Fischeretal}{Fischer et~al.\ }

\begin{document}
\title{\Large Translating Between\\Wavelet Tree and Wavelet Matrix Construction}
\author{Patrick Dinklage\thanks{Technische Universit{\"a}t Dortmund, Department of Computer Science, \href{mailto:patrick.dinklage@tu-dortmund.de}{patrick.dinklage@tu-dortmund.de}}}
\date{}
\maketitle

\begin{abstract}
The wavelet tree (Grossi et al. [SODA, 2003]) and wavelet matrix
(Claude et al. [Inf.~Syst., 2015]) are compact data structures with many
applications such as text indexing or computational geometry.
By continuing the recent research of Fischer et al. [ALENEX, 2018],
we explore the similarities and differences of these heavily related data
structures with focus on their construction.
We develop a data structure to modify construction algorithms for either the
wavelet tree or matrix to construct instead the other.
This modification is \emph{efficient}, in that it does not worsen the
asymptotic time and space requirements of any known wavelet tree or wavelet
matrix construction algorithm.
\end{abstract}

\section{Introduction}
The wavelet tree \cite{DBLP:conf/soda/GrossiGV03} is a data structure with
numerous applications in text indexing, data compression,
computational geometry (as an alternative to fractional cascading) and other
areas \cite{DBLP:journals/iandc/FerraginaGM09,DBLP:journals/jda/Navarro14,
DBLP:conf/latin/MakinenN06}.
Common queries that the wavelet tree can answer efficiently are
\emph{rank} and \emph{select} for any symbol that occurs in the underlying
text, as well as \emph{access} queries to restore said text.
The wavelet matrix \cite{DBLP:journals/is/ClaudeNP15} is a related data
structure with the same asymptotic running times for these queries
However, they are faster in practice, because they requires less
subqueries on bit vectors to be answered.

Both data structures are based on storing $n\lsceil$ bits for the text
of length $n$ over an alphabet of size $\sigma$ and answer access, rank
and select queries in asymptotic time $\lO(\log \sigma)$.
Since they can be used to restore the text via access queries, they can
be seen as different encodings of it.
They differ
\begin{enumerate*}[(a)]
\item in the order these bits are stored, and
\item in the auxiliary data required to answer the queries.
\end{enumerate*}
However, there are many similarities between these two data structures
and it is natural to ask how far these similarities go.
In this work, we focus on the construction process of the data structures.

\paragraph{Related work.}
\Fischeretal \cite{DBLP:conf/alenex/0001KL18} recently showed that there is
a data structure that can be used to efficiently transform any construction
algorithm for the wavelet tree to construct instead the wavelet matrix
without worsening the asymptotic construction times.
This makes it possible to apply techniques used by parallel wavelet tree
construction algorithms, which make use of the tree structure, to the wavelet
matrix, which discards the tree structure.
Their data structure occupies $\lO(n+\sigma\log n)$ bits of space and can be
constructed in time $\lO(n+\sigma)$ using $o(n+\sigma)$ bits of memory.

\paragraph{Our contributions.}
However, they left open whether there is a data structure for the inverse
direction, \ie, whether there is an efficient way to construct
the wavelet tree using a construction algorithm for the wavelet matrix.
In order to learn more about the similarities and differences, we propose
a first solution to this problem by giving the corresponding
data structure of the same asymptotic space requirements as that in
\cite{DBLP:conf/alenex/0001KL18}.
However, our data structure has some limitations in where it can be used,
giving us insights on the differences regarding contained information in
the wavelet tree and matrix.

\begin{table}[t]
\begin{subfigure}[b]{0.45\textwidth}
\centering
\begin{tabular}{c c c c}
$i$ & $(i)_{\setB,2}$ & $((i)_{\setB,2})^R$ & $\text{bitrev}_2(i)$ \\
\midrule
0 & 00 & 00 & 0 \\
1 & 01 & 10 & 2 \\
2 & 10 & 01 & 1 \\
3 & 11 & 11 & 3 \\
\end{tabular}
\label{fig:bit-reversal-2}
\caption{Bit-reversal permutation for $k=2$.}
\end{subfigure}
\begin{subfigure}[b]{0.45\textwidth}
\centering
\begin{tabular}{c c c c}
$i$ & $(i)_{\setB,3}$ & $((i)_{\setB,3})^R$ & $\text{bitrev}_3(i)$ \\
\midrule
0 & 000 & 000 & 0 \\
1 & 001 & 100 & 4 \\
2 & 010 & 010 & 2 \\
3 & 011 & 110 & 6 \\
4 & 100 & 001 & 1 \\
5 & 101 & 101 & 5 \\
6 & 110 & 011 & 3 \\
7 & 111 & 111 & 7 \\
\end{tabular}
\label{fig:bit-reversal-3}
\caption{Bit-reversal permutation for $k=3$.}
\end{subfigure}
\caption{Breakdowns of the bit-reversal permutations for $k=2$ (left) and $k=3$
(right). The first column contains the integers $i < 2^k$, the second shows
their $k$-bit binary representations, the third shows the reversals and the
final column contains the $k$-bit reversal of $i$.}
\label{tab:bit-reversal}
\end{table}

\section{Preliminaries}
Let $T \in \Sigma^n$ be a text over an alphabet $\Sigma$.
For some integer $i<n$, let $T[i]$ be the $i$-th symbol of $T$.
We use zero-based indexing, so that $T[0]$ is the first symbol of $T$ and
$T[n-1]$ is the last.

\paragraph{Computational model.}
For our analysis, we use the \emph{word RAM} model, where we assume that
we can perform arithmetic operations on words of width $\lO(\log n)$ bits in
time $\lO(1)$.

\paragraph{Histogram.}
The \emph{histogram} $H: c \mapsto \occ_T(c)$ of $T$ maps each symbol
$c \in \Sigma$ to its number $\occ_T(c)$ of occurrences in $T$.
The set of those $\sigma$ symbols with $\occ_T(c) > 0$ are the
\emph{effective alphabet} of $T$.
We represent it as the interval $\Sigma' = \ropenint{0}{\sigma}$, so
that the lexicographically smallest symbol is represented by $0$ and the
largest symbol by $\sigma-1$.
Let $\eff_T(c) \in \Sigma'$ be the rank of $c$ in the effective alphabet.
In the \emph{effective transformation} $T'$ of $T$, we set
$T'[i] \coloneqq \eff_T(T[i])$ for each $i<n$.
As an example, consider the
text and alphabet in \autoref{fig:wt-example}. The effective transformation
of the text is $T' = 6\,0\,5\,1\,2\,1\,4\,4\,3\,1\,1$.

\paragraph{C array.}
For every $x \in \Sigma'$, the \emph{$C$ array} contains the accumulated
number of occurrences of symbols in $T'$ that are lexicographically smaller
than $x$.
Formally, it is $C[x] \coloneqq \sum\nolimits_{k=0}^{x-1} \occ_{T'}(k)$.
We furthermore define $C[\sigma] \coloneqq n$.

\paragraph{Bit vectors.}
A \emph{bit vector} is a text over the binary alphabet $\setB = \{0, 1\}$.
Let $B = \setB^n$ be a bit vector of length $n$.
For every position $i < n$, the function $\rank_1(B, i)$ returns the number of
1-bits in $B$ from its beginning up to (including) position $i$.
For a $k > 0$, the function $\select_1(B, k)$ returns the position of the $k$-th
1-bit in $B$.
The functions $\rank_0$ and $\select_0$ are defined analogously for 0-bits.
There is a data structure that can answer rank and select queries for a
fixed $B$ and any $i$ or $k$, respectively, in time $\lO(1)$, requires
$\lo(n)$ bits of memory and can be constructed in time $\lO(n)$
\cite{DBLP:conf/focs/Jacobson89}.

\paragraph{Bit reversal.}
Let $B \in \setB^*$ be a bit vector and let $(B)_\setN \in \setN$ denote the
integer that $B$ is the binary representation of.
For $k > 0$ and an integer $i < 2^k$, we call $(i)_{\setB,k} \in \setB^k$
the $k$-bit binary representation of $i$.
Let $B^R$ denote the reversal of $B$.
We define the \emph{$k$-bit reversal}
$\text{bitrev}_k(i) \coloneqq (((i)_{\setB,k})^R)_\setN$ as the integer
represented by the reversal of $i$'s $k$-bit binary representation.
For a fixed $k$, the \emph{bit-reversal permutation} maps each integer
$i < 2^k$ to its $k$-bit reversal. To give examples, \autoref{tab:bit-reversal}
shows the bit-reversal permutations for $k=2$ and $k=3$.

\subsection{The Wavelet Tree}
The \emph{wavelet tree} \cite{DBLP:conf/soda/GrossiGV03} is a binary
tree of height $\lsceil$ where each node $v$ represents an
interval $[a,b] \subseteq \Sigma'$ of the effective alphabet and is labeled by
a bit vector $B_v \in \setB^+$. $B_v$ contains one bit for each text
position $i$, in text order, where $T'[i] \in [a,b]$: a 0-bit if
$T'[i] \leq \lfloor\frac{a+b}{2}\rfloor$, \ie, if the symbol $T'[i]$
lies in the left half of the represented interval, or a 1-bit otherwise.

\begin{figure}[tb]
\begin{subfigure}{0.7\textwidth}
    \centering
    \includegraphics{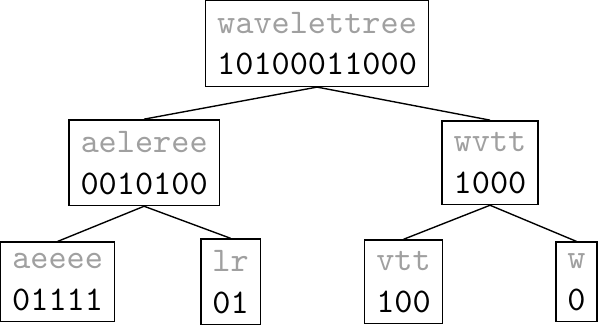}
\end{subfigure}
\begin{subfigure}{0.2\textwidth}
    \parbox[t]{\linewidth}{
    \begin{scriptsize}
    $\Sigma = \{ \texttt{a}, \texttt{e}, \texttt{l}, \texttt{r}, \texttt{t},
    \texttt{v}, \texttt{w} \}\\\Sigma'=[0,7)$
    
    \vspace{1.5ex}
    \renewcommand*{\arraystretch}{0.85}
    \begin{tabular}{l l}
    $c$ & $\eff_T(c)$\\
    \midrule
    \texttt{a} & $0 ~=~ 000_b$\\
    \texttt{e} & $1 ~=~ 001_b$\\
    \texttt{l} & $2 ~=~ 010_b$\\
    \texttt{r} & $3 ~=~ 011_b$\\
    \texttt{t} & $4 ~=~ 100_b$\\
    \texttt{v} & $5 ~=~ 101_b$\\
    \texttt{w} & $6 ~=~ 110_b$\\
    \end{tabular}
    \end{scriptsize}
    }
\end{subfigure}
\caption{
The wavelet tree (left), alphabet, effective alphabet and binary
representations of symbols (right) for $T = \texttt{wavelettree}$.
The texts above the node bit vectors are shown only for comprehensibility;
they are not a part of the node labels and are not stored.
}\label{fig:wt-example}
\end{figure}

The root node represents the entire effective alphabet $\Sigma'$ and thus its
bit vector has length $n$.
A node $v$ has two children iff $a < b$. We apply the described structure
recursively for the left child to represent the interval
$[a, \lfloor\frac{a+b}{2}\rfloor]$ (the \emph{left half}) and the right child to
represent $[\lfloor\frac{a+b}{2}\rfloor+1, b]$ (the \emph{right half}).
Following that, the tree's leaves are those nodes that represent an interval of
size one, \ie, precisely one symbol from the input alphabet ($a = b$).
Since the bit vector of a leaf contains only zero-bits, we need not store
level $\lsceil+1$ of the wavelet tree, because it would consist of leaves only.
\autoref{fig:wt-example} shows an example of a wavelet tree.

The size of any node in the wavelet tree, \ie, the length of its bit vector
label, can be precomputed using the $C$ array:
\begin{observation}\label{obs:node-sizes}
Let $[a,b] \subseteq \Sigma'$ be the alphabet interval represented by a
wavelet tree node $v$.
The length of the bit vector $B_v$ that labels $v$ is $|B_v| = C[b+1] - C[a]$.
\end{observation}

For storing the wavelet tree, we consider the \emph{pointerless}
representation (also known as the \emph{levelwise} representation),
where we concatenate the bit vectors on each level and enhance
them by constant-time rank/select support.
This is is enough information to be able to navigate in the tree
\cite{DBLP:journals/jda/Navarro14}.
The concatenation of bit vectors on any level has a length of at most $n$ bits,
so that the wavelet tree's bit vectors consume at most $n\lsceil$ bits in total.

\subsection{The Wavelet Matrix}
\begin{figure}[tb]
\centering
\includegraphics{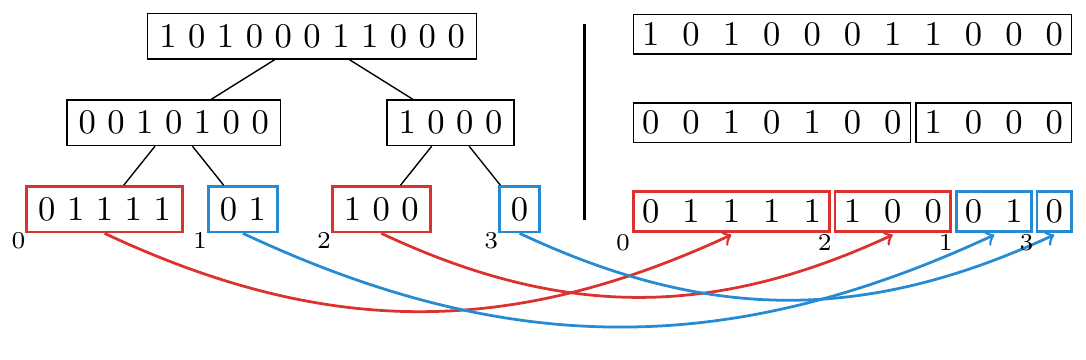}
\caption{Comparison of the node ordering in the wavelet tree (left) and the
wavelet matrix (right). Due to the nature of the bit reversal permutation, the
ordering on the first two levels remains the same in the wavelet matrix. On the
third level, we observe how nodes 0 and 2 (left children of their respective
parents) go to the left part of the corresponding wavelet matrix bit vector
and nodes 1 and 3 (right children of their respective parents) go to the
right.}\label{fig:wm-example}
\end{figure}

The \emph{wavelet matrix} \cite{DBLP:journals/is/ClaudeNP15} can be thought
of as an alternative representation of the pointerless wavelet tree.
In the wavelet tree, in order to retrieve the bit vector
$B^T_\ell$ for level $\ell$, we concatenate the bit vectors of the single nodes
on that level from \emph{left to right}.
In the wavelet matrix, the nodes are
concatenated in a different order to obtain bit vector $B^M_\ell$: all left
children of their respective parents are moved to the left and all right
children are moved to the right.
Like in the pointerless wavelet tree, we concatenate the bit vectors of all
nodes on every level.
\autoref{fig:wm-example} shows an example.
The re-ordering of nodes corresponds to the bit-reversal permutation of the
node ranks on the respective level \cite{DBLP:conf/alenex/0001KL18}.

A practical consequence of the different ordering is that navigation in the
wavelet matrix becomes easier than in the wavelet tree.
In the tree, we need to keep track of the current node's interval
--- its left and right boundary ---
within the respective level's bit vector while navigating.
This can be done using two rank queries on the respective bit vector when
navigating from a node to either child.
In the matrix, the simpler structure makes it feasible to precompute the
left boundary for the right children on each level, all of which have
been concatenated in the right part of the level's bit vector.
This boundary is often referred to as value $z$ in literature, as it
corresponds to the number of zero bits in the bit vector.
We can store $z$ for all levels using negligible $\lO(\log\sigma\log n)$ bits
and use it to save one rank query on each level while navigating.

One could precompute the same information for the wavelet tree.
However, this would require us to store the left boundary of every node,
resulting in $\lO(\sigma \log n)$ bits as there are $\lO(\sigma)$ nodes.
For this reason, the wavelet matrix can be considered more relevant for
practical applications.

\section{Wavelet Tree and Wavelet Matrix Construction}
We continue the research of
\Fischeretal \cite{DBLP:conf/alenex/0001KL18}
and are interested in how a construction algorithm for the wavelet tree
or matrix can be modified efficiently to construct the other.
We consider such a modification \emph{efficient} if the asymptotic time and
space boundaries of the modified construction algorithm are not worsened.
\Fischeretal show that there is a data structure that can
be used to efficiently transform any construction algorithm for the wavelet
tree to construct instead the wavelet matrix.
We propose a data structure for the inverse direction, transforming a wavelet
matrix construction algorithm to one for the wavelet tree, with the same
asymptotic space requirements.

Formally, let us consider the situation where, during the construction of the
wavelet tree, the $i$-th bit is set in bit vector $B^T_\ell$ of level $\ell$ of
(assuming, without loss of generality, the pointerless representation).
\Fischeretal \cite{DBLP:conf/alenex/0001KL18} present
a data structure to efficiently compute a function
$f: (\ell, i) \mapsto (\ell, j)$ so that $j$ is the corresponding position for
the bit to be set in bit vector $B^M_\ell$ of the wavelet matrix.
That is, by modifying the wavelet tree constructor to set the bit at position
$f(\ell, i)$ instead of $i$ on level $\ell$, it instead constructs the wavelet
matrix.
Because $f$ can be computed in constant time, there is no asymptotic overhead.
For input length $n$ and alphabet size $\sigma$, their data structure occupies
${n+\sigma+(\sigma+2)\lceil\log n\rceil}$ bits of space and can be constructed
in time $\lO(n+\sigma)$ using $o(n+\sigma)$ bits of memory, not worsening
the asymptotic construction time and space requirements for any known
wavelet tree constructor.

In the following, we first observe various properties of the wavelet tree that
lead to a similar result for $f$ as that of \cite{DBLP:conf/alenex/0001KL18}.
Based on these observations, we develop a novel data structure for the inverse
$f^{-1}$, which maps $(\ell, j)$ back to $(\ell, i)$ with the same asymptotic
time and space boundaries as for $f$.

\subsection{Locating Nodes and Bit Offsets}\label{sec:locate}

The re-ordering of nodes between the wavelet tree and matrix can be described
by the bit-reversal permutation.
This knowledge makes it easy to translate a node \emph{ID} (the node's rank in
a breadth-first traversal of the tree) between the two data structures.
Based on that, we employ the following strategy
to find data structures for functions $f$ and $f^{-1}$:
given the level and position of the bit to be written, we attempt to find
\begin{enumerate}[(1)]
\item the ID of the node that the bit belongs to, and
\item the position of the node's first bit in its level's bit vector.
\end{enumerate}
With this information available, $f$ and $f^{-1}$ are easy to compute in
constant time.

\paragraph{Bottom level node sizes.}
\autoref{obs:node-sizes} shows the relation
between the $C$ array and the sizes of the wavelet tree's nodes.
This relation is especially interesting regarding the \emph{virtual} bottom-most
level $h=\lsceil$ of a full binary wavelet tree.
We call this level virtual, because all bits on it would be zero and there is
no need to actually store it.
On this level, each node corresponds to a single symbol from the effective
alphabet.
Let node $v_c$ on level $h$ correspond to symbol $c \in \Sigma'$.
We have $|B_{v_c}| = C[c+1] - C[c] = \occ_{T_\eff}(c)$, \ie, the
size of $v_c$ matches the number of occurrences of $c$.

This property is only valid if the wavelet tree is a \emph{full} binary
tree: if it was not, there would be leaves on level $h-1$ and not all nodes on
level $h$ would exist.
Without loss of generality, let us assume from now on that
$\sigma = 2^h$ for some integral $h>0$, \ie, that the alphabet size is a power
of two. Then, the wavelet tree is a full binary tree.
In case $\sigma$ is not a power of two, we introduce artificial symbols that
never occur in the input and are lexicographically \emph{larger} than all
symbols of $\Sigma'$. This way, the empty nodes for these symbols are moved to
the far right of the wavelet tree and can be ignored in the following.

\paragraph{Locating in the wavelet tree.}
We consider the situation where a wavelet tree constructor sets the $i$-th of
bit vector $B^T_\ell$. Let $v(\ell, i)$ be the rank of the wavelet tree node
on level $\ell$ to which the $i$-th bit belongs.
We represent $v(\ell, i)$ relative to the number of the first node on level
$\ell$, \ie, $v(\ell, 0) = 0$ and $v(\ell, n-1) = 2^\ell - 1$.
This representation requires $\ell$ bits, because there are precisely
$2^\ell-1$ nodes on level $\ell$.
Furthermore, let $p(\ell, v)$ be the position of the first bit in $B^T_\ell$
that belongs to node $v$ and let
$\delta_v(\ell, i) \coloneqq i - p(\ell, v(\ell, i))$
be the distance of $i$ from that position.

We take a closer look at $v$ and $p$ on the virtual level $h$ and observe that
\begin{align*}
v(h, i) = \min\{ x ~\vert~ C[x] > i \} - 1.
\end{align*}
This is because each node on this
level corresponds to precisely one symbol from the input alphabet and the $C$
array encodes, for every $c$, the number of symbols in the input that are
lexicographically smaller than $c$. This corresponds to the accumulated sizes
of the node's left siblings. An example of this relation can be seen comparing
\autoref{fig:c-array-ext} and \autoref{fig:wt-nodesizes-sizes}
(in row $\ell=3$).
The node that $i$ belongs to on level $h$ is left of the first node whose
accumulated size --- its entry in the $C$ array --- exceeds $i$.
We can immediately conclude that the first bit that belongs to node $v$ is
located at position
\begin{align*}
p(h, v) = C[v].
\end{align*}

\begin{figure}[tb]
\centering
\begin{tabular}{ r | c c c c c c c c c }
~
&\texttt{\color{greyA0}a}
&\texttt{\color{greyA0}e}
&\texttt{\color{greyA0}l}
&\texttt{\color{greyA0}r}
&\texttt{\color{greyA0}t}
&\texttt{\color{greyA0}v}
&\texttt{\color{greyA0}w}
&{\color{greyA0}$\top$} & ~ \\
$c$ & 0 & 1 & 2 & 3  & 4 & 5 & 6 & 7 & ~ \\
\midrule
$\occ_T(c)$ & 1 & 4 & 1 & 1 & 2 & 1 & 1  & 0  & ~\\
$C[c]$      & 0 & 1 & 5 & 6 & 7 & 9 & 10 & 11 & 11 \\
\end{tabular}
\caption{The histogram and the $C$ array for $T=\texttt{wavelettree}$.
We added the artificial symbol $\top$ so $\sigma=8$ is a power of two. The new
symbol never occurs in $T$ and is lexicographically larger than the other
symbols.}
\label{fig:c-array-ext}
\end{figure}

\begin{figure}[b]
\begin{subfigure}[t]{0.45\textwidth}
\centering
\includegraphics{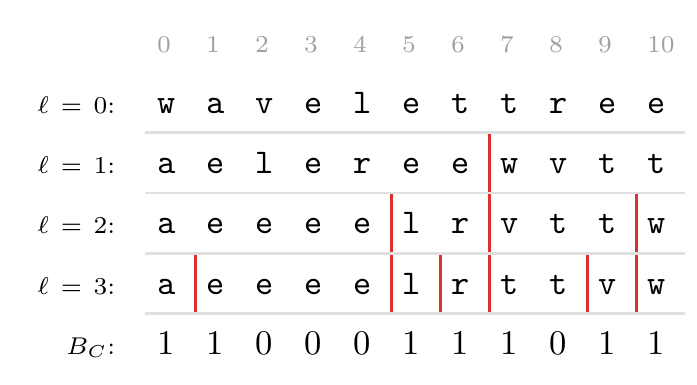}
\caption{The text re-ordering on each level and the bit vector $B_C$.
The vertical lines mark the boundaries of the wavelet tree's nodes.}
\label{fig:wt-nodesizes-tree}
\end{subfigure}
\hfill
\begin{subfigure}[t]{0.45\textwidth}
\centering
\raisebox{4.5ex}{
\includegraphics{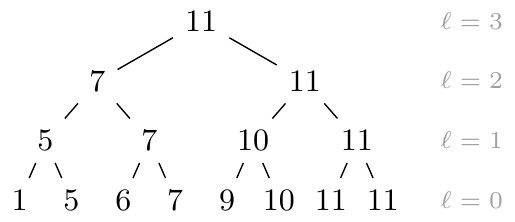}
}
\caption{The accumulated sizes of each of the wavelet tree's nodes. Note that
the rightmost node on the bottom level corresponds to our artificial symbol
$\top$ from \autoref{fig:c-array-ext}.}
\label{fig:wt-nodesizes-sizes}
\end{subfigure}
\caption{Display of the wavelet tree's text re-ordering on each level,
including the virtual level $h=3$, the bit vector $B_C$ and the accumulated
node sizes for our running example text $T = \texttt{wavelettree}$.}
\label{fig:wt-nodesizes}
\end{figure}

How do $v$ and $p$ on level $h$ relate to those on the other levels $\ell < h$
that we are actually interested in? To answer this, we make use of the fact
that our wavelet tree is a full binary tree: the size of a node equals the sum
of its children's sizes, because the children partition the alphabet
interval of their parent.
As a consequence, the \emph{accumulated} size of any node is retained in its
right child, as can be seen in \autoref{fig:wt-nodesizes-sizes}.
Since the $C$ array encodes the accumulated sizes of the nodes on level $h$,
it also implicitly encodes the accumulated sizes of all nodes on levels
$\ell < h$. Following this notion, we can conclude the following relations:
\begin{align*}
v(\ell, i) &=
\left\lfloor \frac{\min\{ x ~\vert~ C[x] > i \}-1}{2^{h-\ell}} \right\rfloor
\end{align*}
and
\begin{align}
p(\ell, v) &= C[v \cdot 2^{h-\ell}] \label{eq:wt-node-offset}.
\end{align}

If the $C$ array is stored in ascending order, the minimum query required to
find $v$ can be answered in time $\lO(\log \sigma)$ using binary search.
However, we seek a computation in constant time. We construct the bit vector
$B_C$ of length $n$ by setting $B_C[k] \coloneqq 1$ if $C[c] = k-1$ for some
$c$ and $B_C[k] \coloneqq 0$ otherwise and prepare it for constant-time rank
queries.
This can be done in time $\lO(n)$ and requires $n + o(n)$ bits of additional
space.
$B_C$ marks the node boundaries on level $h$ of the wavelet tree,
see \autoref{fig:wt-nodesizes-tree} for an example.
We can now compute
\begin{align}
v(\ell, i) &=
\left\lfloor \frac{\text{rank}_1(B_C, i)-1}{2^{h-\ell}} \right\rfloor
\label{eq:wt-node-index}
\end{align}
in constant time.

We now know that the $i$-th bit in $B^T_\ell$ corresponds to the
$(\delta_v)$-th bit in the $v$-th node on level $\ell$ in the wavelet tree.
We can compute $v$, $p$ and $\delta_v$ in constant time using the $C$ array and
rank-enhanced bit vector $B_C$, which together occupy
$\sigma \lceil \log n \rceil + n(1+o(1))$ bits of space.
Asymptotically, this space boundary matches that of the data structure
presented by \Fischeretal \cite{DBLP:conf/alenex/0001KL18}.

\begin{example}
\autoref{fig:wt-nodesizes}, in combination with \autoref{fig:c-array-ext}, shows
an example of the data structure for $T = \texttt{wavelettree}$.
Assume that we are interested in locating the node for bit $i=9$ on level
$\ell=2$.
With \autoref{eq:wt-node-index}, we get
$v(2,9) = \left\lfloor \frac{\text{rank}_1(B_C, 9)-1}{2^{3-2}} \right\rfloor =
\left\lfloor \frac{5}{2} \right\rfloor = 2$.
This means that the bit belongs to the third node on level $2$ (because we start
counting at zero).
Furthermore, with \autoref{eq:wt-node-offset}, we get
$p(2,2) = C[2 \cdot 2^{3-2}] = C[4] = 7$.
This means that the third node on level $2$ starts at position $7$.
Finally, it is $\delta_v(2,9) = 9 - p(2,2) = 9-7 = 2$, so bit $9$ on level
$2$ ultimately corresponds to the third bit of the third node on that level.
\end{example}

\paragraph{Locating in the wavelet matrix.}

The question is how a similar locating can be done for the wavelet matrix.
As described earlier, the bit vector $B^M_\ell$ of the wavelet matrix is
the concatenation of the wavelet tree node bit vectors on level $\ell$ in
bit-reverse order.
To that regard, the wavelet matrix can be represented as a tree just as well
with the nodes re-ordered accordingly.
Even though there are no practical advantages of storing the wavelet matrix as
a tree, this notion will help us find an efficient data structure for
computing $f$ and $f^{-1}$.

We consider the situation where a wavelet matrix constructor sets the $j$-th
bit of bit vector $B^M_\ell$ and are interested in the node to which this bit
belongs. 
Analogously to $v$, $p$ and $\delta_v$, we define
$u(\ell, j)$, $q(\ell, u)$ and
$\delta_u(\ell, j) \coloneqq j - q(\ell, u(\ell, i))$ as the node into which
the written bit belongs, the position of the node's first bit in $B^M_\ell$ and
the distance of $j$ from the node's first bit, respectively.

Due to the re-ordering of the nodes, the correspondences between their
accumulated sizes and the $C$ array, which we observed for the wavelet tree,
are no longer valid for the wavelet matrix.
As a consequence, we need to find a different way to compute $u$ and $q$.

The following observation is useful to find $u$: in both the wavelet
tree and the wavelet matrix \cite[Prop.~1]{DBLP:journals/is/ClaudeNP15},
all occurrences of a symbol $c \in \Sigma'$ belong to the same node on any
level.
Therefore, in order to find the node to which any occurrence of $c$ belongs
on virtual level $h$, it suffices to know to which node the \emph{first}
occurrence of $c$ belongs.
This first occurrence of $c$ on level $h$ is always located at position $C[c]$.
As seen previously, once the node for level $h$ is known, it is easy to narrow
it down to any level $\ell < h$.
Of course, we then have the node in the wavelet \emph{tree}, but in the wavelet
matrix, the nodes are simply permuted in bit-reverse order.
Let $c$ be the symbol from which we computed the bit that we are setting in
$B^M_\ell$. If $c$ is known, we can express
\begin{align}
u(\ell, j, c) &= \text{bitrev}_\ell(v(\ell, C[c])).  \label{eq:wm-node-index}
\end{align}
The consequences of having to know $c$ are discussed later.

\begin{figure}[tb]
\begin{subfigure}[t]{0.45\textwidth}
\centering
\includegraphics{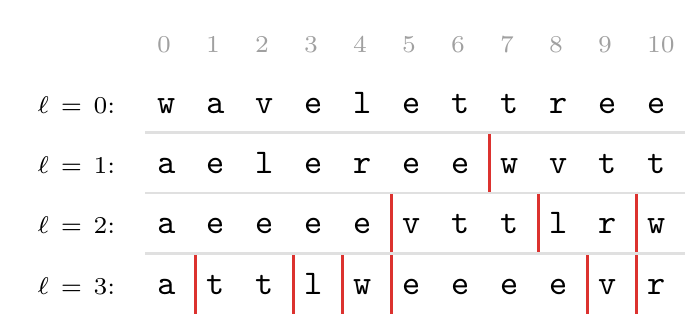}
\caption{The text re-ordering on each level of the wavelet matrix.
The vertical lines mark the boundaries of the \emph{nodes} of the wavelet
matrix.}
\label{fig:wm-nodesizes-tree}
\end{subfigure}
\hfill
\begin{subfigure}[t]{0.45\textwidth}
\centering
\raisebox{0.5ex}{
\includegraphics{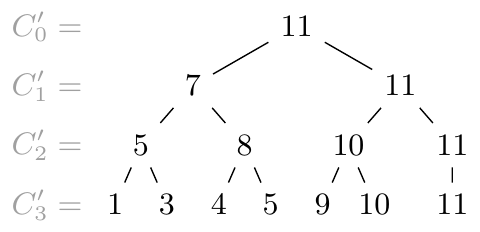}
}
\caption{The accumulated sizes of each of the \emph{nodes} of the wavelet
matrix.
Note that $C_3'$, the bottom level, is not actually needed and depicted
only for the sake of completeness.}
\label{fig:wm-nodesizes-sizes}
\end{subfigure}
\caption{Display of the wavelet matrix's text re-ordering on each level for
running example text $T = \texttt{wavelettree}$.}
\label{fig:wm-nodesizes}
\end{figure}

It remains to compute $q$. As stated above, the $C$ array cannot be used
directly to compute the accumulated node sizes for the wavelet matrix, because
nodes are permuted.
However, the node sizes themselves remain the same and thus, with awareness of
the bit-reversal ordering of nodes on every level, it is easy to precompute the
accumulated node sizes for all nodes of the wavelet matrix using the $C$ array
in time $\lO(\sigma)$. Since we are dealing with a full binary tree of height
$h=\log\sigma$, the accumulated wavelet matrix node sizes can be stored in an
array $C'$ of length $2^h-1 = \sigma - 1$ (since $\sigma$ is a power of two),
occupying $(\sigma-1) \lceil \log n \rceil$ bits of space.
\autoref{fig:wm-nodesizes-sizes} shows an example.
We imagine $C'$ to be a set of arrays $C'_\ell$ for each level $\ell$, so that
the first entry of $C'_\ell$ contains the size of the first node on level $\ell$.
Then, $q$ can be found as follows:
\begin{align}
q(\ell, u) &= \begin{cases}
0            &\text{if } u = 0.\\
C'_\ell[u-1] &\text{if } u > 0.
\end{cases} \label{eq:wm-node-offset}
\end{align}

We then know that the $j$-th bit in $B^M_\ell$ of the wavelet matrix
corresponds to the $\delta_u$-th bit in the $u$-th node's bit vector on level
$\ell$.
We can compute $u$, $q$ and $\delta_u$ in constant time using the arrays $C$
and $C'$ and rank-enhanced bit vector $B_C$, which, in total, occupy
$(2\sigma-1) \lceil \log n \rceil + n(1+o(1))$ bits of space.

\begin{example}
\autoref{fig:wm-nodesizes}, in combination with \autoref{fig:wt-nodesizes}
and \autoref{fig:c-array-ext}, shows an example for the data structure for
$T = \texttt{wavelettree}$.
Assume that we are interested in locating the node for bit $j=9$ on level
$\ell=2$ of the wavelet matrix.
The symbol for which the bit is written is $c=\texttt{r}$ (see
\autoref{fig:wm-nodesizes-tree}).
With \autoref{eq:wm-node-index}, we get
$u(2,9,\texttt{r}) = \text{bitrev}_3(v(2,C[\texttt{r}])) =
\text{bitrev}_3(v(2,6)) = \text{bitrev}_2(1) = 2$.
This means that the bit belongs to the third node on level $2$.
Furthermore, with \autoref{eq:wm-node-offset}, we get $q(2,2) = C'_2[2-1] = 8$.
This means that the third node on level $2$ starts at position $8$.
Finally, it is $\delta_u(2,9) = 9-8 = 1$, so bit $9$ on level $2$ ultimately
corresponds to the second bit of the third node on that level.
\end{example}

\subsection{Translating Between Wavelet Tree and Matrix Construction}
\label{sec:convert-wt-wm}

Using the locating data structures described above, we can express functions
$f$ and $f^{-1}$ as follows:
\begin{align*}
f(\ell, i)
&= q(\ell, \text{bitrev}_\ell(v(\ell, i))) + \delta_v(\ell, i), \\
f^{-1}(\ell, j, c)
&= p(\ell, \text{bitrev}_\ell(u(\ell, j, c))) + \delta_u(\ell, j, c).
\end{align*}

Both $f$ and $f^{-1}$ can be computed in constant time using the arrays $C$,
$C'$ and rank-enhanced bit vector $B_C$.
These occupy
$\sigma \lceil \log n \rceil + (2\sigma-1) \lceil \log n \rceil + n(1+o(1))$
bits of space can be constructed in time $\lO(\sigma+n)$.

\paragraph{Limitations.}
We impose the restriction that for $f^{-1}$, the symbol $c$, for which a bit
is being set in $B^M_\ell$, has to be known when setting the bit.
Even though this bit must ultimately have been computed from $c$, there are
construction algorithms for the wavelet tree that redistribute the bits of
$c$ before constructing the bit vectors
\cite{DBLP:conf/spire/Kaneta18,DBLP:journals/tcs/MunroNV16,
DBLP:conf/soda/BabenkoGKS15,DBLP:conf/cpm/Tischler11}.
Due to the existence of our function $f$ alone, such techniques may as well be
used for the construction of the wavelet matrix.
In this case, $c$ is \emph{not} known when setting the bit in question and
$f^{-1}$ cannot be used.

More generally, in the wavelet tree, $c$ is always implicitly given by the
tree structure itself and implicitly used by $f$ by jumping to the virtual
bottom level to the leaf that would represent $c$ via the $C$ array.
The wavelet matrix discards the tree structure and the information
is lost, so that we need to receive it for from the constructor in order
to compute $f^{-1}$.

\section{Conclusions}
We solved an open theoretical problem concerning the construction of wavelet
trees and wavelet matrices.
We described a data structure that can be used to extend a construction
algorithm for the wavelet matrix to construct instead the wavelet tree with
constant time overhead.
This data structure can be constructed in time $\lO(\sigma+n)$ time and it
requires $\lO(\sigma \log n + n)$ bits of memory, matching the asymptotic
time and space requirements of the data structure described by
\Fischeretal \cite{DBLP:conf/alenex/0001KL18} for
the inverse direction, transforming wavelet tree construction into wavelet
matrix construction.

However, because the wavelet matrix discards wavelet tree's binary tree
structure, we require some additional information from the constructor
for our computations.
This limitation makes our data structure unsuitable for the class of wavelet
matrix constructors that do not keep the entire binary representation of the
input symbols when computing the bit vectors.
To that end, it is still open whether there is a data structure for our
translation function with the same (or lower) asymptotic time and space
requirements that does not require any information other than the position of
the written bit in the wavelet matrix.

\section*{Acknowledgements}
We would like to thank Johannes Fischer and Florian Kurpicz from the
TU Dortmund University's Chair of Algorithm Engineering for the motivation
of this work and the supportive discussions related to the topic.

\bibliographystyle{plainnat}
\bibliography{literature}

\end{document}